\documentclass[ajp,preprint]{revtex4-2}

\usepackage{amsmath}
\usepackage{amsfonts}
\usepackage{amssymb}
\usepackage{graphicx}
\usepackage{enumitem}
\usepackage{dcolumn}
\usepackage{bm}
\usepackage{braket}
\usepackage{ragged2e}
\usepackage{array}
\usepackage{tabularx}
\usepackage{hyperref}
\usepackage{float}

\newcolumntype{Y}{>{\raggedright\arraybackslash}X}
\newcolumntype{P}[1]{>{\raggedright\arraybackslash}p{#1}}

\begin{document}
	\title{From Two-Level Hamiltonians to Quantum Superposition and Measurement: A Traceable Classroom Module}
	
	\author{Boris Kiefer}
	\affiliation{Department of Physics, New Mexico State University, Las Cruces, New Mexico 88003, USA}
	
\begin{abstract}
	We present a traceable instructional module that maps documented learner difficulties in quantum superposition and measurement to explicit learning goals, classroom activities, and assessment evidence. Using the general two-level Hamiltonian as the organizing physics framework, the module implements this mapping through a five-activity classroom sequence and grading rubric targeting four recurrent difficulties: interpreting superposition as physical splitting, confusing a quantum state with its representation or measurement context, mixing theoretical probabilities with finite-sample frequencies, and using inconsistent notation across states, amplitudes, probabilities, and measurement outcomes. The instructional design is organized through a reversible mapping from documented barriers to learning goals, activity responses, and assessment evidence, allowing each assessment item to be traced back to the specific difficulty it is intended to address. The module uses the standard general two-level Hermitian Hamiltonian, represented by a \(2\times2\) matrix, as the physical framework connecting its eigenstates, computational-basis amplitudes, Born probabilities, and sampled outcomes. The contribution is a challenge-targeted instructional design that organizes standard quantum mechanics into a traceable classroom sequence linking documented learner difficulties to learning goals, activities, assessment criteria, and implementation guidance. The mathematical structure uses standard Hamiltonian and measurement conventions, supporting direct incorporation into an undergraduate quantum-mechanics course. 
\end{abstract}
\maketitle
	
\section{Introduction}

Superposition is often one of the first topics in quantum mechanics for which students must coordinate a mathematical state representation with measurement outcomes that have no direct classical analogue. Even in two-state systems, students commonly struggle to distinguish the quantum state from its representation in a chosen basis, interpret amplitudes and probabilities consistently, and connect theoretical probabilities with finite sets of measurement outcomes \cite{Krijtenburg-Lewerissa2017-aj,Singh2015-qu,Marshman2015-um,Bouchee2022-nf}. Reported difficulties include interpreting superposition as physical splitting, making basis- and measurement-context errors, and treating finite-sample fluctuations as evidence that the underlying state or preparation has changed \cite{Marshman2017-hc,Zhu2012-mn,Marshman2017-wg,Hu2023-fn,Borish2024-lh}.

A range of quantum-learning tools already supports undergraduate instruction. Browser-based resources such as PhET and QuVis lower technical barriers through interactive visualizations and broad topic coverage \cite{Perkins2006-fr,kohnle2015-ho}, while platforms such as QuTiP and Qiskit provide flexible numerical modeling in more advanced settings \cite{Johansson2012-uz,Javadi-Abhari2024-bp}. These tools make quantum simulation increasingly accessible, but they do not by themselves determine which documented learner difficulties should be targeted, how those difficulties should be translated into learning goals and activity prompts, or what assessment evidence should demonstrate that the intended reasoning has occurred.

The present module addresses that instructional-design problem through an explicitly traceable mapping from documented learner difficulties to learning goals, activity responses, and assessment evidence. Importantly, the mapping is also intended to be read in reverse: an instructor can begin with an assessment response and trace it back through the corresponding learning goal and activity to the documented difficulty that motivated it. This reversible barrier-to-assessment structure provides both a check on internal alignment and a portable design strategy for adapting the approach to other quantum-mechanics topics.

The module begins with the general Hermitian two-level Hamiltonian
\begin{equation}
	H=h_0I+h_x\sigma_x+h_y\sigma_y+h_z\sigma_z
\end{equation}
and follows the sequence
\begin{equation}
	H
	\longrightarrow
	\{E_\pm,\ket{E_\pm}\}
	\longrightarrow
	\{c_0,c_1\}
	\longrightarrow
	\{|c_0|^2,|c_1|^2\}
	\longrightarrow
	\text{finite measurement statistics}.
\end{equation}
This progression connects a familiar physical object---the Hamiltonian---to its eigensystem, the representation of its eigenstates as superpositions in the computational basis, Born-rule probabilities, and the frequencies obtained from finite measurement samples.

Finite sampling is included as the final step rather than as a separate statistical topic. Students first obtain theoretical probabilities from the eigenstate amplitudes and then compare those probabilities with frequencies generated from a finite number of simulated measurements. This makes explicit that Born-rule probabilities describe the underlying distribution, while experimental or simulated counts fluctuate from one finite realization to another even when the Hamiltonian and quantum state are unchanged \cite{Marshman2017-wg,Borish2024-lh}.

For classroom use, the instructor distributes a single worksheet containing the five activity prompts and associated concept checks. Students alternate between written predictions, interaction with the reference Jupyter simulator, and short explanations. The simulator keeps the Hamiltonian, eigensystem, computational-basis amplitudes, Born probabilities, and sampled frequencies visible in a common interface so that the mathematical sequence remains explicit during the activity. Collected responses can then be graded with the aligned rubric. Supplementary Material provides the concept checks, matched pre/post prompts, answer key, deployment variants, and reference notebook.

The contribution is a portable, traceable instructional architecture that links documented learner difficulties to learning goals, classroom activities, and assessment evidence, with a reference Jupyter notebook, concept checks, rubric, and implementation guidance for a single 50-minute undergraduate quantum-mechanics class meeting. The mathematical structure uses standard Hamiltonian and measurement conventions, allowing direct incorporation into an undergraduate quantum-mechanics course while preserving an explicit trace from each targeted learner difficulty to the evidence used to assess the corresponding reasoning.

\section{Design logic and barrier-to-design mapping}

\subsection{Barrier selection}

The module is organized around four recurrent learner difficulties drawn from the physics-education literature. Barriers were included when they are documented in prior work and can be addressed directly through a two-level Hamiltonian, its eigenstates, and computational-basis measurement. The resulting focus keeps the module suitable for a single undergraduate quantum-mechanics class meeting while connecting the mathematical representation of a two-level system to measurement probabilities and finite-sample outcomes.

The selected barriers are:
\begin{itemize}[leftmargin=*]
	
	\item \textbf{C1:} superposition interpreted as physical splitting rather than as a state represented by amplitudes in a chosen basis \cite{Marshman2017-hc,Passante2015-ll};
	
	\item \textbf{C2:} basis and measurement-context errors, including difficulty distinguishing a quantum state from its representation in a particular basis \cite{Zhu2012-mn,Hu2023-fn};
	
	\item \textbf{C3:} confusion between Born-rule probabilities and finite-sample measurement frequencies \cite{Marshman2017-wg,Borish2024-lh};
	
	\item \textbf{C4:} notation-level ambiguity among Hamiltonians, state amplitudes, basis states, probabilities, and measurement statements \cite{Baily2010-yv,Baily2015-cu,Merzel2024-al,Singh2015-qu}.
	
\end{itemize}

The objective of the present work is to connect a general two-level Hamiltonian explicitly to its eigensystem, represent its eigenstates in the computational basis, obtain Born-rule probabilities, and interpret finite-sample measurement outcomes.

\subsection{Learning goals and backward mapping}

The four barriers motivate five learning goals. After completing the module, students should be able to:

\begin{itemize}[leftmargin=*]
	
	\item \textbf{LG1:} determine how the eigenstates of a two-level Hamiltonian are represented in the computational basis and identify when those eigenstates are superpositions;
	
	\item \textbf{LG2:} relate the parameters $h_0,h_x,h_y,$ and $h_z$ to physically distinct features of the eigensystem, including eigenstate composition, relative phase, common energy shift, and level splitting;
	
	\item \textbf{LG3:} obtain computational-basis measurement probabilities from the amplitudes of an energy eigenstate using the Born rule;
	
	\item \textbf{LG4:} distinguish theoretical probabilities from finite-sample frequencies and explain how increasing the number of trials affects observed fluctuations without changing the underlying state;
	
	\item \textbf{LG5:} communicate the relationships among Hamiltonians, eigenstates, amplitudes, probabilities, and sampled outcomes using consistent quantum-mechanical notation.
	
\end{itemize}

The design follows a backward-mapping structure: documented learner difficulties motivate the learning goals; the learning goals determine the simulator displays and activity prompts; and the prompts determine the assessment evidence. The mapping is intentionally reversible. Each assessment item can be traced back through its associated learning goal and activity response to the documented barrier that motivated it. Table~\ref{tab:per_to_design} makes both directions explicit. This reverse mapping provides a practical check on instructional alignment and a portable design strategy: instructors adapting the framework to another quantum topic can begin either with a documented learner difficulty or with desired assessment evidence and verify that the intervening learning goals and activities remain aligned.

Barrier C4 is treated as a cross-cutting difficulty rather than assigned to a single activity. The simulator deliberately places the Hamiltonian, its eigensystem, computational-basis amplitudes, Born-rule probabilities, and finite-sample frequencies in a common visual sequence. Written prompts then require students to distinguish these quantities explicitly. This repeated alignment reinforces notation and interpretation across the full sequence rather than isolating them in a single exercise.

Because the mapping is defined by the relationship among learner difficulty, learning goal, activity, and assessment evidence rather than by the particular simulator, the same design logic can be adapted to other quantum-mechanics topics by replacing the targeted barriers and associated instructional responses while preserving the alignment structure.

\begin{table}[H]
	\caption{Reversible mapping from documented student difficulties to learning goals, instructional responses, and assessment evidence. Reading from left to right shows the forward design logic; reading from the assessment column back to the barrier shows the reverse trace used to check instructional alignment. CC and PP labels refer to supplementary concept checks and pre/post prompts.}
	\label{tab:per_to_design}
	\scriptsize
	\begin{tabular}{@{}l@{\hspace{0.012\textwidth}}l@{\hspace{0.025\textwidth}}l@{\hspace{0.025\textwidth}}l@{\hspace{0.025\textwidth}}l@{}}
		\hline \\[-0.1cm]
		\parbox[t]{0.06\textwidth}{\textbf{Barrier}} &
		\parbox[t]{0.20\textwidth}{\textbf{PER finding(s)}} &
		\parbox[t]{0.21\textwidth}{\textbf{Learning goal and design response}} &
		\parbox[t]{0.19\textwidth}{\textbf{Where implemented}} &
		\parbox[t]{0.25\textwidth}{\textbf{Assessment evidence and reverse trace \\[0.2cm]}} \\
		\hline \\[-0.1cm]
		
		\parbox[t]{0.06\textwidth}{C1} &
		\parbox[t]{0.20\textwidth}{\justifying\noindent
			Superposition interpreted as physical splitting rather than as a state representation \cite{Marshman2017-hc,Passante2015-ll}.} &
		\parbox[t]{0.21\textwidth}{\justifying\noindent
			\textbf{LG1.} Begin with diagonal Hamiltonians, then introduce off-diagonal coupling and display the resulting eigenstates explicitly as linear combinations of $\ket{0}$ and $\ket{1}$.} &
		\parbox[t]{0.19\textwidth}{\justifying\noindent
			Activities~1 and~2; eigenstate table in the simulator.} &
		\parbox[t]{0.25\textwidth}{\justifying\noindent
			\textbf{CC1, PP1} $\rightarrow$ LG1 $\rightarrow$ C1. Evidence: the student explains when and why an energy eigenstate is a superposition in the computational basis without interpreting superposition as physical splitting.} \\[2pt]
		
		\parbox[t]{0.06\textwidth}{C2} &
		\parbox[t]{0.20\textwidth}{\justifying\noindent
			Basis and measurement-context errors \cite{Zhu2012-mn,Hu2023-fn}.} &
		\parbox[t]{0.21\textwidth}{\justifying\noindent
			\textbf{LG1, LG3.} Display energy eigenstates and their computational-basis components simultaneously and connect each amplitude directly to its Born-rule measurement probability.} &
		\parbox[t]{0.19\textwidth}{\justifying\noindent
			Activities~1--3; eigenstate, amplitude, and probability displays.} &
		\parbox[t]{0.25\textwidth}{\justifying\noindent
			\textbf{CC1/PP1 and CC3/PP3} $\rightarrow$ LG1/LG3 $\rightarrow$ C2. Evidence: the student distinguishes the energy eigenbasis from the computational-basis representation and obtains the corresponding computational-basis probabilities.} \\[2pt]
		
		\parbox[t]{0.06\textwidth}{C3} &
		\parbox[t]{0.20\textwidth}{\justifying\noindent
			Finite-sample measurement frequencies confused with theoretical probabilities \cite{Marshman2017-wg,Borish2024-lh}.} &
		\parbox[t]{0.21\textwidth}{\justifying\noindent
			\textbf{LG4.} Display Born-rule probabilities and finite-shot frequencies together while allowing the number of trials and random realization to vary independently of the Hamiltonian.} &
		\parbox[t]{0.19\textwidth}{\justifying\noindent
			Activity~5; computational-basis sampling panel.} &
		\parbox[t]{0.25\textwidth}{\justifying\noindent
			\textbf{CC4, PP4} $\rightarrow$ LG4 $\rightarrow$ C3. Evidence: the student explains finite-sample variability and its dependence on $N$ without attributing it to a change in the Hamiltonian or quantum state.} \\[2pt]
		
		\parbox[t]{0.06\textwidth}{C4} &
		\parbox[t]{0.20\textwidth}{\justifying\noindent
			Notation and interpretation disconnect among mathematical representations and measurement claims \cite{Baily2010-yv,Baily2015-cu,Merzel2024-al,Singh2015-qu}.} &
		\parbox[t]{0.21\textwidth}{\justifying\noindent
			\textbf{LG2, LG5.} Use conventional notation throughout and distinguish Hamiltonian parameters, eigenstates, amplitudes, probabilities, and sampled frequencies in a fixed visual sequence.} &
		\parbox[t]{0.19\textwidth}{\justifying\noindent
			All activities and simulator output; emphasized explicitly in Activities~3 and~4 and in the assessment rubric.} &
		\parbox[t]{0.25\textwidth}{\justifying\noindent
			\textbf{CC2/PP2 and CC5/PP5} $\rightarrow$ LG2/LG5 $\rightarrow$ C4. Evidence: the student correctly interprets Hamiltonian parameters and produces notation-consistent explanations that distinguish state amplitudes, Born-rule probabilities, and sampled frequencies. \\[-0.1cm]} \\
		
		\hline
	\end{tabular}
\end{table}

\section{Two-level Hamiltonian framework and simulator}

The module is built around the most general Hermitian two-level Hamiltonian,
\begin{equation}
	H=h_0 I+h_x\sigma_x+h_y\sigma_y+h_z\sigma_z
	=
	\begin{pmatrix}
		h_0+h_z & h_x-i h_y\\
		h_x+i h_y & h_0-h_z
	\end{pmatrix},
\end{equation}
where $h_0,h_x,h_y,$ and $h_z$ are real parameters. This form provides a direct connection between the matrix representation of a physical two-level system, its energy eigenstates, and superposition in the computational basis.

It is useful to write
\begin{equation}
	\mathbf{h}=(h_x,h_y,h_z),
	\qquad
	r=|\mathbf{h}|=\sqrt{h_x^2+h_y^2+h_z^2},
\end{equation}
and, for $r>0$,
\begin{equation}
	h_x=r\sin\theta\cos\phi,\qquad
	h_y=r\sin\theta\sin\phi,\qquad
	h_z=r\cos\theta.
\end{equation}
The Cartesian and spherical representations are displayed simultaneously in the simulator so that students can connect changes in the Hamiltonian coefficients to changes in the eigenstates.

The eigenvalues are
\begin{equation}
	E_\pm=h_0\pm r,
\end{equation}
with energy splitting
\begin{equation}
	\Delta E=E_+-E_-=2r.
\end{equation}
The coefficient $h_0$ therefore produces a common energy shift but does not change the energy splitting or the eigenstates. Equivalently, in time evolution it contributes only the global phase factor
\begin{equation}
	e^{-ih_0t/\hbar}.
\end{equation}
This provides a simple example of a Hamiltonian parameter that changes the numerical eigenvalues without changing observable computational-basis probabilities.

For $r>0$, one convenient phase convention for the normalized eigenstates is
\begin{align}
	\ket{E_+}
	&=
	\cos\frac{\theta}{2}\ket{0}
	+
	e^{i\phi}\sin\frac{\theta}{2}\ket{1},
	\\
	\ket{E_-}
	&=
	-e^{-i\phi}\sin\frac{\theta}{2}\ket{0}
	+
	\cos\frac{\theta}{2}\ket{1}.
\end{align}
Thus each energy eigenstate is, in general, a superposition of the computational-basis states,
\begin{equation}
	\ket{E_\pm}=c_0^{(\pm)}\ket{0}+c_1^{(\pm)}\ket{1}.
\end{equation}
A computational-basis measurement gives the Born-rule probabilities
\begin{equation}
	P(k|E_\pm)=\left|c_k^{(\pm)}\right|^2,
	\qquad k\in\{0,1\}.
\end{equation}
The simulator displays the complex amplitudes and their squared magnitudes together so that the connection between state-vector coefficients and measurement probabilities is explicit.

The special case $r=0$ is treated separately. In this limit $H=h_0I$, the two eigenvalues are degenerate and the Hamiltonian does not select a unique eigenbasis. The simulator therefore identifies the degeneracy rather than displaying an arbitrary numerical eigenbasis.

Finite measurement statistics are modeled by repeated computational-basis measurements of either eigenstate. For $N$ trials, the sampled frequencies fluctuate around the Born-rule probabilities and converge toward them as $N$ increases. The simulator displays theoretical probabilities and sampled frequencies together, allowing students to distinguish a probability predicted from the quantum state from the frequency obtained in a finite experimental sample.

\begin{figure}[!t]
	\centering
	
		\textbf{(a)}\\[2mm]
		\includegraphics[width=0.9\textwidth]{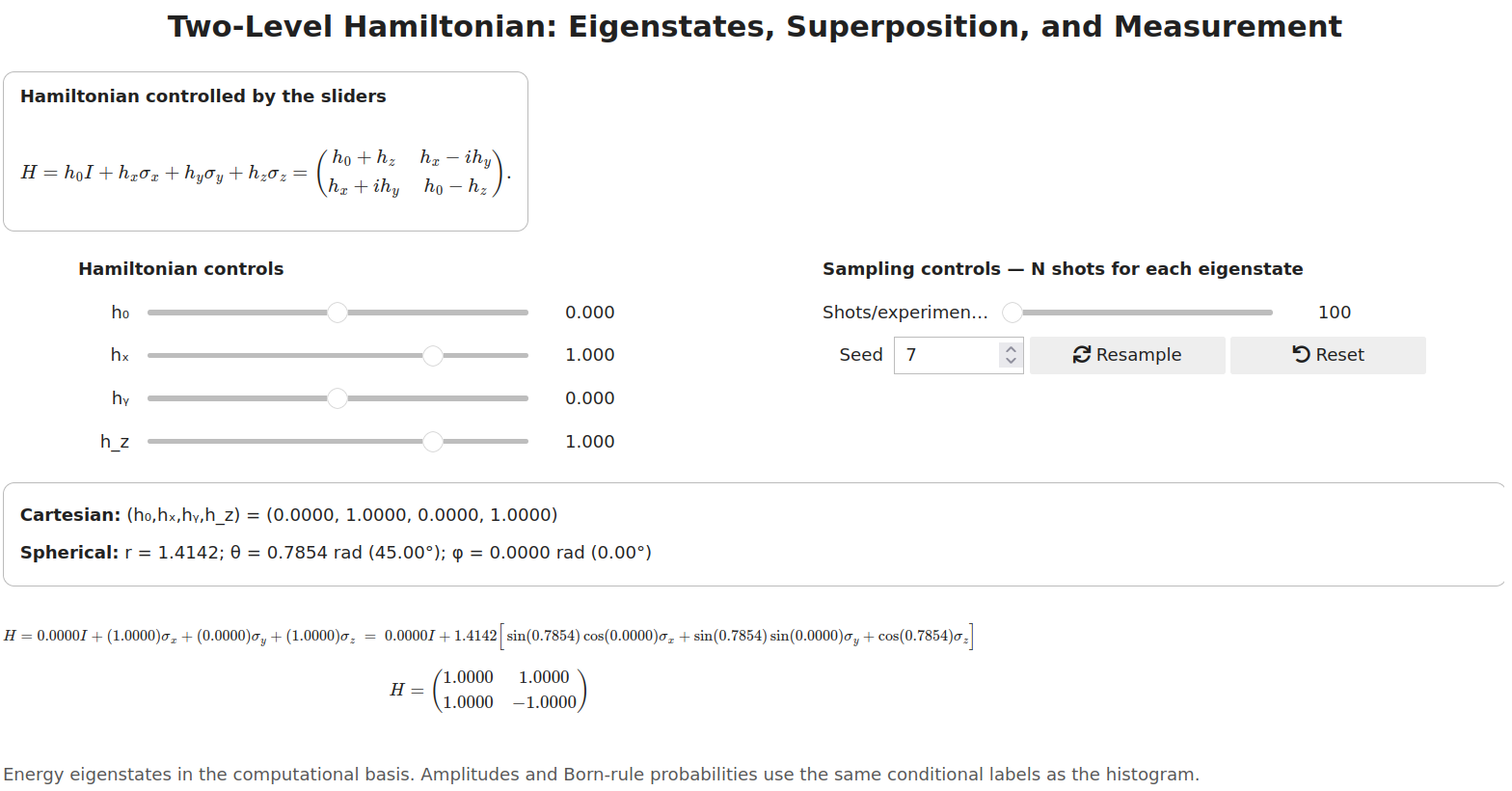}
	
	\vspace{2mm}
	
		\textbf{(b)}\\[2mm]
		\includegraphics[width=0.9\textwidth]{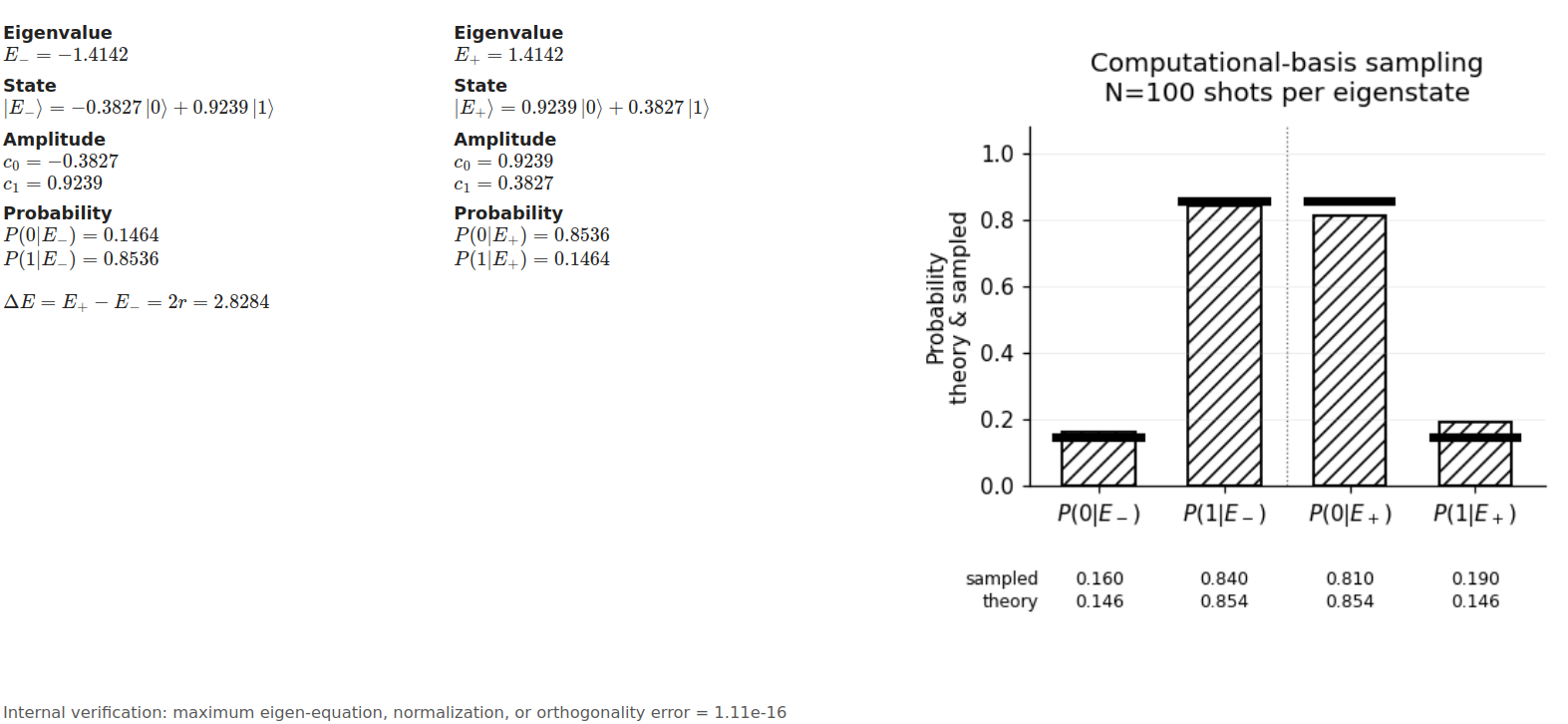}
	
	\caption{Jupyter simulator used in the activity.
		(a) Hamiltonian and sampling controls together with the Cartesian and
		spherical parameterizations and the corresponding Hamiltonian matrix.
		(b) Energy eigenvalues and eigenstates, computational-basis amplitudes
		and Born-rule probabilities, and finite-shot measurement results.
		The sampled frequencies are shown together with the theoretical
		probabilities so that finite-sample fluctuations can be compared
		directly with the Born-rule prediction.}
	\label{fig_sim}
\end{figure}

The included Jupyter notebook is the reference implementation used throughout the activities (Fig.~\ref{fig_sim}). Four sliders control $h_0,h_x,h_y,$ and $h_z$, while the numerical output displays the Cartesian and spherical Hamiltonian parameters, the matrix $H$, its eigenvalues and eigenstates, the computational-basis amplitudes and probabilities, and finite-shot sampling. The interface is intentionally compact so that the mathematical sequence
\begin{equation}
	H
	\longrightarrow
	\{E_\pm,\ket{E_\pm}\}
	\longrightarrow
	\{c_0,c_1\}
	\longrightarrow
	\{|c_0|^2,|c_1|^2\}
	\longrightarrow
	\text{finite sampling}
\end{equation}
remains visible during the activity.

\section{Activity sequence}

The module is designed for a single class meeting, with optional extensions assigned outside class. The five activities follow the same progression as the simulator: Hamiltonian structure $\rightarrow$ eigenstates $\rightarrow$ computational-basis amplitudes $\rightarrow$ Born-rule probabilities $\rightarrow$ finite measurement statistics. Together, the sequence addresses LG1--LG5.

Each activity uses a predict--observe--explain structure. Students first record a written prediction, then compare it with the simulator output, and finally write a brief explanation reconciling the two. Predictions are completed individually before simulator interaction, and the final written explanations provide the principal assessment artifacts. This structure is consistent with evidence that recording predictions before observation can improve conceptual engagement and reduce uncritical acceptance of confirming outcomes \cite{crouch2001peer,kohnle2015-ho}.

\paragraph*{Activity 1: diagonal Hamiltonians and computational-basis eigenstates.}
Students begin with $h_x=h_y=0$ and $h_z\neq0$. The Hamiltonian is diagonal in the computational basis, so its energy eigenstates are $\ket{0}$ and $\ket{1}$ up to phase and ordering. Before using the simulator, students predict the two eigenvalues, identify the corresponding eigenstates, and predict the computational-basis measurement probabilities. They then vary the magnitude and sign of $h_z$ and explain which quantities change and which do not.

The activity establishes the basic Hamiltonian--state connection and introduces the physical role of $h_z$. It also establishes the notation used throughout the module by distinguishing the Hamiltonian, its energy eigenstates, and the computational-basis states.

\paragraph*{Activity 2: off-diagonal coupling and superposition.}
Students next set $h_z=0$, $h_y=0$, and choose $h_x\neq0$. The Hamiltonian is no longer diagonal in the computational basis, and its energy eigenstates become equal-weight superpositions of $\ket{0}$ and $\ket{1}$. Students predict whether a computational-basis measurement of either energy eigenstate will be deterministic or probabilistic before examining the displayed eigenvectors.

They then identify the complex amplitudes $c_0$ and $c_1$ in
\begin{equation}
	\ket{E_\pm}
	=
	c_0^{(\pm)}\ket{0}
	+
	c_1^{(\pm)}\ket{1}
\end{equation}
and use their squared magnitudes to predict $P(0|E_\pm)$ and $P(1|E_\pm)$. The activity makes explicit that an energy eigenstate can appear as a superposition when represented in a different basis, while the Born rule converts the corresponding amplitudes into measurement probabilities.

\paragraph*{Representative in-class prompt.}
To make the classroom implementation concrete, the following prompt is used during Activity~2 after students have examined the diagonal case in Activity~1:

\begin{quote}
	Set $h_0=h_y=h_z=0$ and choose $h_x=1$.
	
	\emph{Predict:} Before running the simulator, write the two energy eigenstates in the computational basis. Will a measurement of either eigenstate in the computational basis give a definite outcome or probabilistic outcomes? Explain your prediction.
	
	\emph{Observe:} Compare your prediction with the displayed eigenstates, amplitudes, and theoretical probabilities. Record $P(0|E_-)$, $P(1|E_-)$, $P(0|E_+)$, and $P(1|E_+)$.
	
	\emph{Explain:} In 2--3 sentences, explain why the energy eigenstates are superpositions of $\ket{0}$ and $\ket{1}$ even though each is a definite energy state. Your explanation should distinguish the energy eigenbasis from the computational basis and should not describe the system as physically splitting between two states.
\end{quote}

For this setting,
\begin{equation}
	H=\sigma_x=
	\begin{pmatrix}
		0 & 1\\
		1 & 0
	\end{pmatrix},
\end{equation}
with eigenstates, up to overall phase and labeling convention,
\begin{equation}
	\ket{E_\pm}
	=
	\frac{\ket{0}\pm\ket{1}}{\sqrt{2}}.
\end{equation}
Thus each energy eigenstate is a definite state of the Hamiltonian while simultaneously being an equal-amplitude superposition in the computational basis, giving
\begin{equation}
	P(0|E_\pm)=P(1|E_\pm)=\frac{1}{2}.
\end{equation}
The intended conceptual conclusion is that superposition does not mean that the physical system has split into two parts. It means that, relative to a specified basis, the quantum state is represented by more than one nonzero complex amplitude, whose squared magnitudes determine the corresponding measurement probabilities.

\begin{figure}[!t]
	\centering
	\includegraphics[width=0.95\columnwidth]{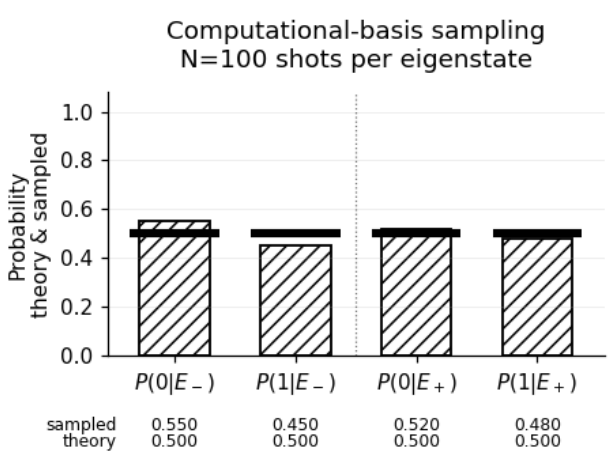}
	\caption{Representative simulator output for Activity~2 with $h_0=h_y=h_z=0$ and $h_x=1$, so that $H=\sigma_x$. The energy eigenstates are equal-amplitude superpositions of the computational-basis states, giving $P(0|E_\pm)=P(1|E_\pm)=1/2$. The finite-shot bars illustrate the sampled frequencies obtained from the same theoretical probabilities.}
	\label{fig:sigmax_example}
\end{figure}

\paragraph*{Activity 3: general two-level Hamiltonian.}
Students vary $h_x,h_y,$ and $h_z$ and compare the Cartesian Hamiltonian coefficients with the spherical parameters $r,\theta,$ and $\phi$. For several Hamiltonians they record the eigenvalues, eigenstate coefficients, and computational-basis probabilities.

The prompts direct attention to three relationships. First, the direction of $\mathbf{h}=(h_x,h_y,h_z)$ determines the eigenstates, while its magnitude $r$ determines the level splitting. Second, the polar angle $\theta$ controls the relative magnitudes of the two computational-basis components. Third, a nonzero $h_y$ generally introduces complex relative phases through $\phi$. Students identify which changes affect the complex amplitudes and which affect their squared magnitudes.

This activity connects a general Hermitian $2\times2$ Hamiltonian to its eigenstates, interprets those eigenstates as superpositions in the computational basis, and obtains the corresponding measurement probabilities.

\paragraph*{Activity 4: common energy shifts and invariant state properties.}
Students hold $h_x,h_y,$ and $h_z$ fixed while varying $h_0$. Before using the simulator, they predict the effect on the eigenvalues, energy splitting, eigenvectors, and computational-basis probabilities. They then observe that both eigenvalues shift by the same amount, while
\begin{equation}
	\Delta E=E_+-E_-=2r
\end{equation}
and the eigenstates and computational-basis probabilities remain unchanged.

Students explain this behavior from the identity contribution $h_0I$. As an optional extension, the result can also be connected to
\begin{equation}
	e^{-iHt/\hbar}
	=
	e^{-ih_0t/\hbar}
	e^{-i(\mathbf{h}\cdot\boldsymbol{\sigma})t/\hbar},
\end{equation}
showing that $h_0$ contributes only a global phase to the time evolution. The activity isolates the physically distinct roles of $h_0$ and $\mathbf{h}$ and therefore directly requires students to distinguish energy shifts, state vectors, and observable probabilities in a notation-consistent explanation.

\paragraph*{Activity 5: finite sampling and measurement statistics.}
Students choose a Hamiltonian for which neither computational-basis probability is zero or one and hold the Hamiltonian fixed while varying only the number of measurement trials $N$. They first record the theoretical probabilities $P(k|E_\pm)$ obtained from the eigenstate amplitudes and then compare these values with the sampled frequencies produced by the simulator.

Repeating the acquisition with different random seeds shows that the sampled frequencies fluctuate even though the Hamiltonian, eigenstates, and theoretical probabilities are unchanged. Students explain why finite-sample frequencies need not equal the Born-rule probabilities for small $N$, why the typical discrepancy decreases as $N$ increases, and why this variation does not imply a change in the quantum state.

If desired, the expected scale of the fluctuations can be estimated from the binomial standard error
\begin{equation}
	SE(\hat p)
	=
	\sqrt{\frac{p(1-p)}{N}},
\end{equation}
where $p$ is the Born-rule probability. This final activity links the theoretical probability obtained from the state amplitudes to experimentally sampled outcomes, requiring students to distinguish explicitly among amplitudes, probabilities, and observed frequencies.

\subsection{Assessment design}

Assessment is built at two levels: brief in-class concept checks (CC1--CC5) and matched pre/post prompts (PP1--PP5). All items are designed for paper collection and rubric-based grading without requiring a learning management system or automated scoring. Table~\ref{tab:rubric} summarizes the rubric, while the full item text and answer keys appear in Supplementary Material S5 and S6, respectively.

The rubric is designed for single-grader use in a typical course context. Each performance level is defined by the presence or absence of a specific identifiable element rather than by holistic impression. The criteria follow the instructional sequence from Hamiltonian interpretation through eigenstate superposition, Born-rule probabilities, and finite-sample measurement statistics. The barrier labels provide the reverse trace to Table~\ref{tab:per_to_design}, so that each scored learning goal remains connected to the documented learner difficulty that motivated it.

\begin{table*}[t]
	\caption{Grading rubric for the module (10 points total). Barrier labels provide the reverse trace from assessment evidence to the documented learner difficulties in Table~\ref{tab:per_to_design}.}
	\label{tab:rubric}
	\begin{tabular}{@{}ll@{}}
		\hline
		\textbf{Criterion} & \textbf{Performance levels} \\
		\hline
		
		\parbox[t]{1.35in}{LG1 (C1, C2)\\0--2 pts} &
		\parbox[t]{5.4in}{\raggedright
			Hamiltonian--state connection: correctly identifies when an energy eigenstate is a superposition in the computational basis and relates this to the Hamiltonian structure (2); partially correct connection (1); incorrect (0).} \\[4pt]
		
		\parbox[t]{1.35in}{LG2 (C4)\\0--2 pts} &
		\parbox[t]{5.4in}{\raggedright
			Hamiltonian-parameter interpretation: correctly distinguishes the roles of $h_0$ and $\mathbf{h}=(h_x,h_y,h_z)$ in common energy shift, level splitting, eigenstate composition, and relative phase (2); partially correct (1); incorrect (0).} \\[4pt]
		
		\parbox[t]{1.35in}{LG3 (C2)\\0--2 pts} &
		\parbox[t]{5.4in}{\raggedright
			Probability prediction: correctly obtains computational-basis probabilities from the squared magnitudes of the state amplitudes and verifies normalization (2); correct setup with minor error (1); incorrect (0).} \\[4pt]
		
		\parbox[t]{1.35in}{LG4 (C3)\\0--2 pts} &
		\parbox[t]{5.4in}{\raggedright
			Sampling interpretation: correctly distinguishes finite-sample frequency fluctuations from changes in the Hamiltonian or quantum state and explains the effect of increasing the number of trials (2); partly correct (1); incorrect (0).} \\[4pt]
		
		\parbox[t]{1.35in}{LG5 (C4)\\0--2 pts} &
		\parbox[t]{5.4in}{\raggedright
			Notation and interpretation: consistently distinguishes Hamiltonian parameters, energy eigenstates, computational-basis states, amplitudes, probabilities, and sampled frequencies in symbolic and verbal explanations (2); minor inconsistency (1); incorrect or substantially inconsistent (0).} \\
		
		\hline
	\end{tabular}
\end{table*}

Representative prompts illustrate the alignment between assessment and activity sequence. For LG1, students explain why introducing an off-diagonal Hamiltonian term can produce energy eigenstates that are superpositions of $\ket{0}$ and $\ket{1}$. For LG2, students predict the effect of changing $h_0$ while holding $h_x,h_y,$ and $h_z$ fixed and identify which displayed quantities change and which remain invariant. For LG3, students use the displayed coefficients
\begin{equation}
	\ket{E_\pm}
	=
	c_0^{(\pm)}\ket{0}
	+
	c_1^{(\pm)}\ket{1}
\end{equation}
to determine $P(0|E_\pm)$ and $P(1|E_\pm)$. For LG4, students compare theoretical probabilities with finite-shot frequencies and explain why increasing $N$ reduces the typical discrepancy without changing the underlying state. LG5 is reinforced throughout through notation-consistent written explanations.

\section{Instructional contribution and scope}

The paper contributes a compact instructional progression that connects several ideas often encountered separately:
\begin{equation}
	H
	\longrightarrow
	\{E_\pm,\ket{E_\pm}\}
	\longrightarrow
	\{c_0,c_1\}
	\longrightarrow
	\{|c_0|^2,|c_1|^2\}
	\longrightarrow
	\text{finite measurement statistics}.
\end{equation}
The general two-level Hamiltonian provides a standard physical starting point, while the simulator keeps its Cartesian and spherical representations, eigensystem, computational-basis amplitudes, Born-rule probabilities, and finite-shot frequencies visible in a single interface.

The contribution is a traceable instructional architecture that uses standard two-level quantum mechanics to connect documented learner difficulties with explicit learning goals, classroom activities, assessment evidence, and implementation guidance. The mathematical structure is intentionally conventional so that instructors can focus on the conceptual progression rather than first introducing a simulator-specific state parameterization or measurement convention.
The simulator and supplementary materials serve as implementation components of this instructional design rather than as the primary contribution.

The present work reports the design, instructional alignment, and reference implementation of the module; it does not report measured learning gains or classroom efficacy. The included assessment artifacts are intended to support subsequent evaluation of student reasoning and module usability.

The module deliberately focuses on the static eigensystem of a general two-level Hamiltonian and computational-basis measurement of its energy eigenstates. It does not treat density matrices, mixed states, continuous-variable systems, multi-qubit entanglement, sequential projective measurement with explicit state update, or general time-dependent evolution. Although the identity contribution $h_0I$ is connected briefly to the global phase factor $e^{-ih_0t/\hbar}$, unitary time evolution is not developed as a separate instructional topic. These extensions require additional formalism and are better treated in subsequent modules rather than within a single class meeting.

\section{Conclusion}

We have presented a classroom-oriented module that uses the general two-level Hamiltonian as a direct route from a physical model to quantum superposition and measurement statistics. Students begin with diagonal Hamiltonians whose eigenstates coincide with the computational basis, introduce off-diagonal terms that generate superposition states, explore the general complex eigenvectors of a Hermitian $2\times2$ Hamiltonian, distinguish common energy shifts from level splitting, and finally compare Born-rule probabilities with finite-shot measurement frequencies.

The central instructional sequence is deliberately compact:
\begin{equation}
	H
	\longrightarrow
	\text{eigenstates}
	\longrightarrow
	\text{superposition amplitudes}
	\longrightarrow
	\text{probabilities}
	\longrightarrow
	\text{sampled outcomes}.
\end{equation}

The accompanying Jupyter notebook displays each step simultaneously and uses standard Hamiltonian and computational-basis notation throughout. The main contribution is the aligned instructional design built around this progression, in which documented learner difficulties are connected explicitly to classroom activities, assessment prompts, grading criteria, and implementation materials suitable for a single undergraduate quantum-mechanics class meeting.

By beginning with the Hamiltonian, the module keeps the mathematical representation tied to a familiar physical object while retaining access to general two-level eigenstates. The finite-sampling activity then connects the ideal Born-rule prediction to the statistical character of actual measurement records. Together, these elements provide a compact bridge between introductory state-vector notation and the interpretation of experimentally observed quantum probabilities.

\section*{Data and code availability}
The reference Jupyter simulator, README, and software requirements are publicly available at
\url{https://github.com/boriskiefer/two-level-superposition-module}.

\section*{author declarations}
The author has no conflict to disclose.
	
\bibliographystyle{unsrt}
\bibliography{sim_superposition_ajp}

\clearpage

\section*{Supplementary Material}
\subsection*{S1. Deployment modes summary}

Three deployment modes are supported. In the \emph{full} 50-minute mode, instructors use all five activities in sequence as described in the main text. Activities~1--4 develop the connection from Hamiltonian structure to eigenstates, computational-basis amplitudes, and Born-rule probabilities; Activity~5 connects those probabilities to finite measurement statistics.

In a \emph{compressed} mode, Activities~1--3 are completed in class and Activities~4 and~5 are assigned as follow-up work. This preserves the Hamiltonian-to-superposition progression as the in-class core while retaining the common-energy-shift and finite-sampling activities for subsequent reinforcement.

In a \emph{flipped} mode, Activity~1 and the introductory portion of Activity~2 are completed before class using the reference notebook and worksheet. Class time can then focus on the general Hamiltonian in Activity~3, the interpretation of $h_0$ in Activity~4, and the finite-sampling comparison in Activity~5.

The learning goals remain unchanged across deployment modes. Instructors using a compressed or flipped implementation can select the concept checks associated with the activities completed during class and use the remaining prompts as homework or follow-up assessment.

\subsection*{S2. General two-level Hamiltonian and eigensystem}

The reference simulator uses the general Hermitian two-level Hamiltonian
\begin{equation}
	H
	=
	h_0I+h_x\sigma_x+h_y\sigma_y+h_z\sigma_z
	=
	\begin{pmatrix}
		h_0+h_z & h_x-i h_y\\
		h_x+i h_y & h_0-h_z
	\end{pmatrix},
\end{equation}
where all four coefficients are real.

Define
\begin{equation}
	r=\sqrt{h_x^2+h_y^2+h_z^2}.
\end{equation}
For $r>0$, the Cartesian coefficients may be written as
\begin{align}
	h_x &= r\sin\theta\cos\phi,\\
	h_y &= r\sin\theta\sin\phi,\\
	h_z &= r\cos\theta,
\end{align}
with
\begin{equation}
	\theta
	=
	\operatorname{atan2}
	\left(
	\sqrt{h_x^2+h_y^2},h_z
	\right),
	\qquad
	\phi
	=
	\operatorname{atan2}(h_y,h_x).
\end{equation}

The eigenvalues are
\begin{equation}
	E_\pm=h_0\pm r,
\end{equation}
and the level splitting is
\begin{equation}
	\Delta E=E_+-E_-=2r.
\end{equation}

One convenient phase convention for the normalized eigenstates is
\begin{align}
	\ket{E_+}
	&=
	\cos\left(\frac{\theta}{2}\right)\ket{0}
	+
	e^{i\phi}
	\sin\left(\frac{\theta}{2}\right)\ket{1},
	\\
	\ket{E_-}
	&=
	-e^{-i\phi}
	\sin\left(\frac{\theta}{2}\right)\ket{0}
	+
	\cos\left(\frac{\theta}{2}\right)\ket{1}.
\end{align}

Thus,
\begin{equation}
	\ket{E_\pm}
	=
	c_0^{(\pm)}\ket{0}
	+
	c_1^{(\pm)}\ket{1},
\end{equation}
with computational-basis probabilities
\begin{equation}
	P(k|E_\pm)
	=
	\left|c_k^{(\pm)}\right|^2,
	\qquad
	k\in\{0,1\}.
\end{equation}

For $r=0$, the Hamiltonian reduces to
\begin{equation}
	H=h_0I.
\end{equation}
The two eigenvalues are then degenerate and the Hamiltonian does not select a unique eigenbasis. The reference simulator reports this degeneracy rather than displaying an arbitrary numerical eigenbasis.

\subsection*{S3. Common energy shifts and global phase}

The parameter $h_0$ shifts both energy eigenvalues by the same amount but does not alter the eigenvectors or the energy splitting. This follows directly from
\begin{equation}
	H=h_0I+\mathbf{h}\cdot\boldsymbol{\sigma},
\end{equation}
because adding a multiple of the identity to a matrix leaves its eigenvectors unchanged.

If time evolution is introduced as an optional extension,
\begin{equation}
	U(t)
	=
	e^{-iHt/\hbar}
	=
	e^{-ih_0t/\hbar}
	e^{-i(\mathbf{h}\cdot\boldsymbol{\sigma})t/\hbar}.
\end{equation}
The factor $e^{-ih_0t/\hbar}$ multiplies the entire state by a common phase. Therefore changing $h_0$ alone leaves the computational-basis probabilities unchanged.

This result is used in Activity~4 to distinguish a common energy shift from changes in level splitting or eigenstate composition. For the core activity, the global-phase relation provides the needed connection between $h_0$ and the invariance of the computational-basis probabilities.

\subsection*{S4. Standard error for binary outcomes}

Let $X_i\in\{0,1\}$ denote the $i$th computational-basis measurement, with
\begin{equation}
	\mathbb{P}(X_i=1)=p,
\end{equation}
and let
\begin{equation}
	\hat{p}
	=
	\frac{1}{N}\sum_{i=1}^{N}X_i.
\end{equation}
For independent trials,
\begin{equation}
	\mathrm{Var}(\hat{p})
	=
	\frac{p(1-p)}{N},
\end{equation}
so that
\begin{equation}
	SE(\hat{p})
	=
	\sqrt{\frac{p(1-p)}{N}}.
\end{equation}

In a classroom calculation, $p$ may be replaced by the theoretical Born-rule probability displayed by the simulator. The $N^{-1/2}$ dependence provides the central result needed for Activity~5 and CC4/PP4: increasing the number of measurements reduces the characteristic finite-sample fluctuation without altering either the Hamiltonian or the quantum state.

\subsection*{S5. Concept checks and pre/post prompts}

The concept checks and matched pre/post prompts follow the five learning goals defined in the main text. Their labels provide the assessment end of the reversible mapping in Table~I of the main manuscript.

\medskip
\noindent\textbf{CC1 (LG1).}
Consider
\begin{equation}
	H=h_z\sigma_z
\end{equation}
with $h_z\neq0$. The energy eigenstates coincide with the computational-basis states. Now set $h_z=0$ and use
\begin{equation}
	H=h_x\sigma_x
\end{equation}
with $h_x\neq0$. How do the energy eigenstates change when expressed in the computational basis? Explain why the eigenstates of the second Hamiltonian are represented as superpositions of $\ket{0}$ and $\ket{1}$ in that basis.

\medskip
\noindent\textbf{CC2 (LG2).}
Suppose $h_x$, $h_y$, and $h_z$ are held fixed while $h_0$ is increased. Predict what happens to:
\begin{enumerate}[label=(\alph*)]
	\item $E_+$ and $E_-$,
	\item the level splitting $\Delta E$,
	\item the eigenstates, and
	\item the computational-basis probabilities.
\end{enumerate}
Briefly justify your answer.

\medskip
\noindent\textbf{CC3 (LG3).}
An energy eigenstate is
\begin{equation}
	\ket{E}
	=
	\frac{\sqrt{3}}{2}\ket{0}
	+
	\frac{i}{2}\ket{1}.
\end{equation}
Determine $P(0|E)$ and $P(1|E)$ and verify that the probabilities sum to one. Does the factor $i$ in the second amplitude itself represent a probability?

\medskip
\noindent\textbf{CC4 (LG4).}
For a fixed Hamiltonian and eigenstate, suppose the Born-rule probability for outcome $\ket{1}$ is $p=0.40$. One simulated acquisition uses $N=100$ trials and another uses $N=1000$ trials. Compare the expected standard errors. Does changing $N$ change the Hamiltonian, eigenstate, or theoretical probability?

\medskip
\noindent\textbf{CC5 (LG5).}
A student writes:
\begin{quote}
	``The coefficient of $\ket{1}$ is $0.36$, so the amplitude is $0.36$ and the probability is also $0.36$.''
\end{quote}
Identify the ambiguity or error in this statement. Rewrite it so that the amplitude and probability are distinguished correctly.

\bigskip
\noindent\textbf{PP1 (LG1).}
In 2--4 sentences, explain how an energy eigenstate can be a superposition when written in the computational basis. Your explanation should distinguish the Hamiltonian eigenbasis from the basis used to represent or measure the state.

\medskip
\noindent\textbf{PP2 (LG2).}
For an instructor-selected set of values $(h_0,h_x,h_y,h_z)$, identify which Hamiltonian parameters control the common energy shift, level splitting, eigenstate component magnitudes, and relative phase. Explain your reasoning using the simulator output.

\medskip
\noindent\textbf{PP3 (LG3).}
For an instructor-selected energy eigenstate
\begin{equation}
	\ket{E}
	=
	c_0\ket{0}
	+
	c_1\ket{1},
\end{equation}
calculate the computational-basis probabilities and state explicitly how the complex amplitudes are converted into probabilities.

\medskip
\noindent\textbf{PP4 (LG4).}
Two acquisitions use the same Hamiltonian and energy eigenstate but different numbers of trials. Explain why the measured frequencies can differ, how the expected fluctuation scales with $N$, and why the difference does not imply that the quantum state changed.

\medskip
\noindent\textbf{PP5 (LG5).}
A student writes:
\begin{quote}
	``The Hamiltonian coefficient $h_x=0.6$ is the amplitude for outcome $\ket{1}$, so $P(1)=0.6$. In 100 measurements we should therefore obtain outcome $\ket{1}$ exactly 60 times.''
\end{quote}
Diagnose and correct this explanation. Your response should distinguish Hamiltonian parameters, state amplitudes, Born-rule probabilities, and finite-sample frequencies.

\subsection*{S6. Answer key and scoring guidance}

\noindent\textbf{CC1 / PP1 (LG1).}
For $H=h_z\sigma_z$, the computational basis diagonalizes the Hamiltonian, so the energy eigenstates are $\ket{0}$ and $\ket{1}$ up to phase and ordering. For $H=h_x\sigma_x$,
\begin{equation}
	\ket{E_\pm}
	=
	\frac{1}{\sqrt{2}}
	\left(
	\ket{0}\pm\ket{1}
	\right)
\end{equation}
up to phase and eigenvalue ordering. Full credit requires recognizing that ``superposition'' refers to the representation of an eigenstate in the computational basis; it does not imply physical splitting of a single system.

\medskip
\noindent\textbf{CC2 / PP2 (LG2).}
Changing $h_0$ by an amount $\delta$ gives
\begin{equation}
	E_\pm\rightarrow E_\pm+\delta.
\end{equation}
Both eigenvalues therefore shift by the same amount, while
\begin{equation}
	\Delta E=2r
\end{equation}
is unchanged. The eigenvectors and computational-basis probabilities are also unchanged. More generally, $r=|\mathbf{h}|$ determines the level splitting, while the direction of $\mathbf{h}$ determines the eigenstate composition. In the spherical parameterization, the polar angle $\theta$ controls the relative magnitudes of the computational-basis components, while the azimuthal angle $\phi$ controls their relative complex phase in the chosen eigenstate convention.

\medskip
\noindent\textbf{CC3 / PP3 (LG3).}
For
\begin{equation}
	\ket{E}
	=
	\frac{\sqrt{3}}{2}\ket{0}
	+
	\frac{i}{2}\ket{1},
\end{equation}
the probabilities are
\begin{equation}
	P(0|E)
	=
	\left|\frac{\sqrt{3}}{2}\right|^2
	=
	\frac{3}{4},
\end{equation}
and
\begin{equation}
	P(1|E)
	=
	\left|\frac{i}{2}\right|^2
	=
	\frac{1}{4}.
\end{equation}
They sum to one. Full credit requires distinguishing the complex amplitude $i/2$ from its squared magnitude $1/4$.

\medskip
\noindent\textbf{CC4 / PP4 (LG4).}
For $p=0.40$,
\begin{equation}
	SE_{100}
	=
	\sqrt{\frac{0.4(0.6)}{100}}
	\approx0.049,
\end{equation}
while
\begin{equation}
	SE_{1000}
	=
	\sqrt{\frac{0.4(0.6)}{1000}}
	\approx0.0155.
\end{equation}
The uncertainty decreases by approximately $\sqrt{10}$ when $N$ increases from 100 to 1000. Changing $N$ does not change the Hamiltonian, eigenstate, or theoretical probability; it changes only the statistical precision of the finite sample.

\medskip
\noindent\textbf{CC5 / PP5 (LG5).}
A probability is the squared magnitude of an amplitude, not generally the amplitude itself. For example, if
\begin{equation}
	c_1=0.6,
\end{equation}
then
\begin{equation}
	P(1)=|c_1|^2=0.36.
\end{equation}
A correct explanation must distinguish the Hamiltonian parameters that define the model, the amplitudes that define the state representation, the Born-rule probabilities obtained from their squared magnitudes, and the finite-sample frequencies obtained from repeated measurements.

For PP5, $h_x$ is a Hamiltonian parameter rather than a state amplitude. The state amplitudes are obtained from the corresponding eigenstate, Born-rule probabilities are their squared magnitudes, and finite-sample frequencies fluctuate around those probabilities rather than being fixed exactly by them.
	
\end{document}